\documentclass[%
 reprint, 
 amsmath,amssymb,
 aps,
 prl,
 floatfix,
]{revtex4-2}

\usepackage{graphicx}
\usepackage{dcolumn}
\usepackage{bm}
\usepackage{afterpage}

\begin{document}

\preprint{APS/PRL}

\title{Direct Measurement of Inertial Impact and Propulsive Force in a Eukaryotic Swimmer}

\author{Katsuya Shimabukuro}
 \email{kshimabu@ube-k.ac.jp}
\affiliation{Department of Chemical and Biological Engineering, National Institute of Technology (KOSEN) Ube College, Yamaguchi, 755-8555 Japan}
\author{Kosaku Horinaga}
\affiliation{Department of Chemical and Biological Engineering, National Institute of Technology (KOSEN) Ube College, Yamaguchi, 755-8555 Japan}
\author{Kazumo Wakabayashi}
\affiliation{Department of Chemical and Biological Engineering, National Institute of Technology (KOSEN) Ube College, Yamaguchi, 755-8555 Japan}
\author{Hikaru Emoto}
\affiliation{Department of Chemical and Biological Engineering, National Institute of Technology (KOSEN) Ube College, Yamaguchi, 755-8555 Japan}
\author{Noriko Ueki}
\affiliation{Science Research Center, Hosei University, Tokyo, 102-8160, Japan}
\author{Ken-ichi Wakabayashi}
\affiliation{Department of Industrial Life Sciences, Faculty of Life Sciences, Kyoto Sangyo University, Kyoto, 603-8555, Japan}
\author{Noriyo Mitome}
 \email{mitome@sz.tokoha.ac.jp}
\affiliation{Department of Chemical and Biological Engineering, National Institute of Technology (KOSEN) Ube College, Yamaguchi, 755-8555 Japan}
\affiliation{Faculty of Education, Tokoha University, Shizuoka, Japan}

\date{\today}

\begin{abstract}
The transduction of force into motion for microswimmers at intermediate Reynolds numbers (Re $\sim$ 1), where inertia becomes relevant, is a fundamental problem in active matter. Using the multicellular alga \textit{Volvox} as a model physical system, we perform the first direct measurements that deconvolve a swimmer's inertial impact force from its motor's propulsive force. We discover a $\sim$30 Hz propulsive pulse, the mechanical signature of collective ciliary action. This high-frequency motor output drives a fluctuating velocity in the low-Re \textit{V. carteri}, but is mechanically filtered by the inertia of the larger \textit{V. ferrisii}, resulting in a smooth swimming trajectory. Our work demonstrates that for swimmers beyond the Stokes regime, kinematics are not a direct proxy for the underlying motor dynamics, a foundational assumption in the study of microscopic motility.
\end{abstract}

\maketitle

The transition from viscous-dominated (Re $\ll$ 1) to inertia-influenced (Re $\sim$ 1) dynamics is a critical, yet poorly understood, regime in the physics of active matter. In the Stokes regime, an organism's kinematics are a direct proxy for its motor's force output \cite{Purcell1977}. As swimmers increase in size, this assumption may break down. While previous methods could measure quasi-static stall forces, the dynamic, inertial components of motility have remained inaccessible. The transient impact force generated by the momentum of a swimming organism during deceleration has, to our knowledge, never been directly measured. Here, we use two species of the multicellular alga \textit{Volvox} as a tunable physical system to probe this regime. By measuring the force from \textit{V. carteri} (Vc, diameter $L \approx 200~\mu$m) and the larger \textit{V. ferrisii} (Vf, $L \approx 800~\mu$m), we directly test how organismal inertia filters the output of a complex biological motor. Using typical swimming speeds of $v \approx 150~\mu$m/s for Vc \cite{Drescher2009} and $v \approx 1$~mm/s for large volvocine algae \cite{Goldstein2015}, we confirm that Vc operates in the low-Re regime (Re $\approx 0.03$) while Vf operates in the intermediate-Re regime (Re $\approx 0.8$), where inertia is significant (see Supplemental Material [4] for calculation).

\begin{figure}[b!]
\includegraphics[width=\columnwidth]{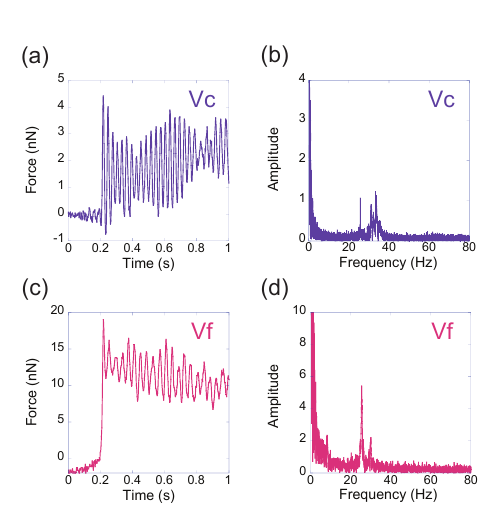}
\caption{\label{fig:force_pulse} Force traces measured with an optical lever system during collision reveal a $\sim$30 Hz propulsive oscillation. (a) Representative force trace for \textit{V. carteri} (Vc). (b) Power spectrum of the trace in (a). (c) Representative force trace for \textit{V. ferrisii} (Vf). (d) Power spectrum of the trace in (c).}
\end{figure}

    \begin{figure}[t!]
    \includegraphics[width=\columnwidth]{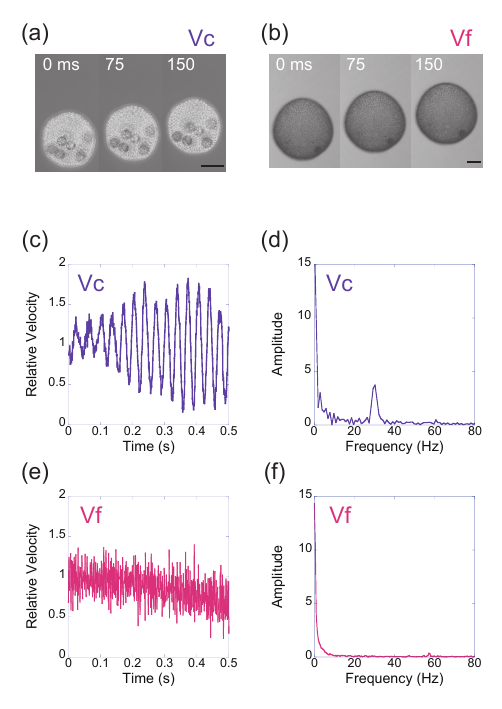}
    \caption{\label{fig:kinematics} Free-swimming kinematics, measured by high-speed microscopy, show inertial damping of the force pulse. (a) Time-lapse images of Vc swimming. (b) Time-lapse images of Vf swimming. Scale bars, 100 $\mu$m. (c) The velocity of Vc oscillates strongly, corresponding to (d) a distinct peak at $\sim$30 Hz in its power spectrum. (e) In contrast, the velocity of Vf is nearly constant, and (f) the $\sim$30 Hz peak in its power spectrum is strongly attenuated.}
    \end{figure}

We measured the propulsive force using a validated optical lever system (see Supplemental Material [4]). Analysis of collision events revealed that the force is not steady but oscillates periodically (Fig.~\ref{fig:force_pulse}). For Vc, the force trace shows a strong, regular oscillation [Fig.~\ref{fig:force_pulse}(a)], whose power spectrum has a dominant peak at $\sim$30 Hz [Fig.~\ref{fig:force_pulse}(b)]. Remarkably, Vf exhibits a propulsive force oscillating at a nearly identical frequency [Fig.~\ref{fig:force_pulse}(c,d)]. This frequency is in excellent agreement with the $\sim$32 Hz ciliary beat frequency that drives the metachronal waves in Vc \cite{Brumley2012}. The oscillation in the total organismal force is a direct consequence of these waves: because a finite number of wavelengths fit on the spheroid's surface, the vector sum of the thousands of individual ciliary force vectors is not perfectly constant in time. Our measurement is therefore the first mechanical signature of this imperfect spatiotemporal averaging. While the frequency is conserved, the force from Vc is highly pulsatile (mean $F_{\text{min}}/F_{\text{max}} = 0.05$), whereas the force from Vf is nearly continuous (mean $F_{\text{min}}/F_{\text{max}} = 0.73$), a result of averaging over a larger number of cells ($\sim$5000 in Vf compared to $\sim$2000 in Vc; see Supplemental Material [4]).

This same motor signature drives dramatically different free-swimming kinematics, measured via high-speed imaging (Fig.~\ref{fig:kinematics}). The pulsatile force in Vc drives a fluctuating velocity with a corresponding $\sim$30 Hz peak in its power spectrum [Fig.~\ref{fig:kinematics}(c,d)]. In stark contrast, the more massive Vf exhibits a nearly constant swimming velocity, with the 30 Hz peak strongly attenuated in its power spectrum [Fig.~\ref{fig:kinematics}(e,f)]. This demonstrates that the organism's own inertia acts as a mechanical low-pass filter, smoothing the trajectory by damping the high-frequency force fluctuations from the ciliary motor.

\begin{figure}[t!]
\includegraphics[width=\columnwidth]{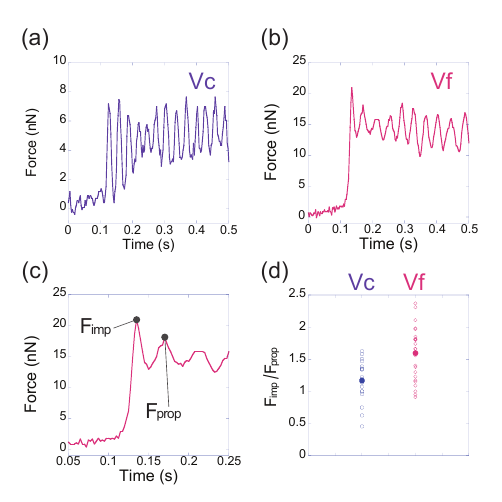}
\caption{\label{fig:impact} Inertial impact force is a signature of Vf motility. (a,b) Representative force traces showing the onset of collision for (a) Vc and (b) Vf. (c) Magnified view of the collision onset in (b), defining the peak impact force ($F_{\text{imp}}$) and the peak of the first subsequent propulsive cycle ($F_{\text{prop}}$). (d) The ratio $F_{\text{imp}}/F_{\text{prop}}$ is $\sim$1 for Vc but significantly $>$1 for Vf.}
\end{figure}

\begin{figure}[t!]
\includegraphics[width=\columnwidth]{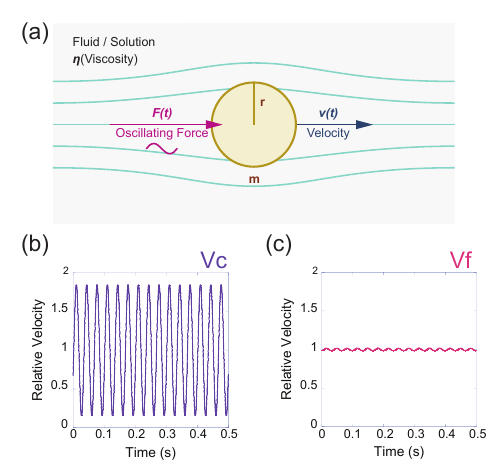}
\caption{\label{fig:simulation} Hydrodynamic simulation confirms inertial damping. (a) Schematic of the 1D model. (b,c) Simulated velocity traces for (b) Vc and (c) Vf, subjected to a 30 Hz force pulse. The model correctly predicts significant velocity oscillations for Vc and a damped, smooth trajectory for Vf.}
\end{figure}
Our dynamic collision assay, in contrast to static stall-force measurements, allows for the first time the direct measurement of the inertial impact force. Figure~\ref{fig:impact} reveals a stark difference in this parameter between the two species. Collisions with Vf are initiated by a large, transient impact force spike ($F_{\text{imp}}$), a direct consequence of its momentum ($p=mv$), which is absent in the low-Re Vc [Fig.~\ref{fig:impact}(a,b)]. The ratio $F_{\text{imp}}/F_{\text{prop}}$, where $F_{\text{prop}}$ is the peak of the first propulsive cycle immediately following the impact, quantitatively captures this effect [Fig.~\ref{fig:impact}(c)]. For Vc, this ratio is $\sim$1, whereas for Vf, $F_{\text{imp}}$ is significantly greater than $F_{\text{prop}}$ [Fig.~\ref{fig:impact}(d)], providing irrefutable mechanical proof of inertia's dominant role.

This inertial filtering is explained by the competition between two timescales: the particle's viscous relaxation time, $\tau_r = m/6\pi\eta r$, and the period of the force, $\tau_f \approx 33$ ms. Their ratio defines the Stokes number, St = $\tau_r/\tau_f$. For Vc, $\tau_r \approx 2.2~\mu$s, giving St $\approx 7 \times 10^{-5}$. For Vf, $\tau_r \approx 36~\mu$s, giving St $\approx 1 \times 10^{-3}$. For St $\ll 1$, velocity is expected to track force. While both values are small, the Stokes number for Vf is over an order of magnitude larger than for Vc. This increased St is sufficient to cause the phase lag and amplitude damping that constitutes the low-pass filtering effect. This is confirmed by a 1D hydrodynamic model, $m(dv/dt) = F(t) - 6\pi\eta rv$, which uses these parameters to correctly capture the fluctuating velocity of Vc and the smooth trajectory of Vf [Fig.~\ref{fig:simulation}(a-c)].

In conclusion, we have provided the first direct measurement of the net propulsive force generated by metachronal waves. We have shown that for microswimmers at intermediate Reynolds numbers, organismal inertia acts as a mechanical low-pass filter, decoupling the high-frequency output of the biological motor from the emergent swimming kinematics. This provides a mechanical basis for understanding how physical constraints, such as inertia, can act as a selection pressure during the evolution of multicellularity \cite{Ueki2024}. Our work establishes that beyond the Stokes regime, an organism's own mass can fundamentally decouple motor dynamics from swimming kinematics, a principle that must be incorporated into future models of active matter.

\begin{acknowledgments}
The authors gratefully acknowledge Professor Kazutaka Fujita for insightful discussions that contributed to this work. This research was funded in part by the Japan
Society for the Promotion of Science (JSPS) KAKENHI,
grant numbers UP23K05725 (K.S.), JP23K05702 (N.M.),21K06295 (N.U.), and 23K22711, 23K23905, 23K18136, and 24H01495 (K.W.). Support for N.M. was also provided by the Cooperative Research Program of the Network Joint Research Center for Materials and Devices. 
\end{acknowledgments}

\newpage

\title{Supplemental Material for: Direct Measurement of Inertial Impact and Propulsive Force in a Eukaryotic Swimmer}

\author{Katsuya Shimabukuro et al.}
\maketitle

\section{Materials and Methods}

\subsection{Strain and Culture}
\textit{Volvox carteri \textnormal{f.} nagariensis} strain Eve 10 (=UTEX1885, $\sim$2000 cells/colony) and \textit{Volvox ferrisii} strain MI01 (=NIES-4029, $\sim$5000 cells/colony) were cultured in Standard Volvox Medium (SVM) under a 16h light/8h dark cycle at 26 $^\circ$C [S1]. Before experiments, mature colonies were collected by filtration through a 40-$\mu$m mesh.

\subsection{Direct Force Measurement and Signal Validation}
The propulsive force of swimming \textit{Volvox} spheroids was measured using a scanning probe microscope (SPM-9700, Shimadzu). This technique monitors the deflection of a microfabricated cantilever by tracking a reflected 670 nm laser spot with a quadrant photodiode [S2]. A probe (CL-RT800PSA-1, Olympus) with two cantilevers of differing spring constants (0.15 N/m and 0.57 N/m) was used. Measurements were performed in a custom glass dish containing 0.5 ml of \textit{Volvox} suspension. The total laser intensity (I), cantilever deflection (D), and torsion (T) signals were recorded at 250 Hz with a digital oscilloscope (PicoScope 4424A, Pico Technology). The deflection signal was converted to force using the manufacturer-provided inverse optical lever sensitivity (InvOLS) and Hooke's Law.

\begin{figure}[h!]
\includegraphics[width=\textwidth]{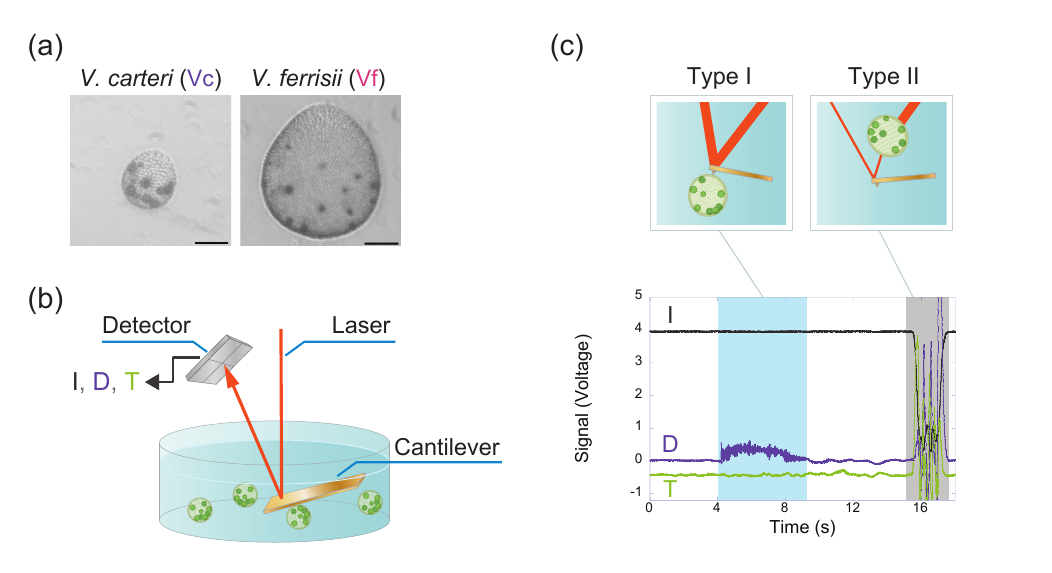}
\caption{\label{fig:setup} \textbf{Experimental setup and signal validation.} (a) Photomicrographs of \textit{V. carteri} (Vc) and \textit{V. ferrisii} (Vf). Scale bars, 100 $\mu$m. (b) Schematic of the optical lever system. (c) Representative raw signals demonstrating the protocol for validating genuine collision events. A genuine collision (Type I, blue shaded region) is characterized by a clean deflection signal (D) with stable total laser intensity (I). An optical artifact (Type II) is identified by a sharp drop in intensity and a large torsional signal (T).}
\end{figure}

As shown in Fig.~\ref{fig:setup}(c), a robust protocol was used to distinguish genuine collision events from optical artifacts. A genuine collision (Type I) is characterized by a clean deflection signal with stable total laser intensity. An optical artifact (Type II), caused by the spheroid obstructing the laser path, is identified by a sharp drop in intensity and a large torsional signal. Only validated Type I events were used for analysis.

\subsection{High-Speed Imaging}
The swimming behavior of Vc and Vf was observed using a high-speed camera (FASTCAM SA5, Photron) mounted on an inverted microscope (IX70, Olympus). Spheroids swam in a chamber formed by two coverslips separated by a 1-mm spacer. Images were acquired at 1000 Hz using a 660 nm LED to minimize phototaxis. Velocity analysis and Fast Fourier Transform (FFT) were performed using ImageJ (NIH) and KaleidaGraph (Synergy Software), respectively.

\subsection{Estimation of Reynolds Number}
The Reynolds number (Re) for a swimming organism is defined as Re = $\rho v L / \mu$, where $\rho$ and $\mu$ are the density and dynamic viscosity of the fluid, respectively, $v$ is the characteristic swimming speed of the organism, and $L$ is its characteristic length (diameter). We assume the properties of water at 25 $^\circ$C ($\rho \approx 1000$~kg/m$^3$ and $\mu \approx 0.001$~Pa$\cdot$s).

For \textit{V. carteri}, with a measured diameter $L \approx 200~\mu$m and a typical swimming speed of $v \approx 150~\mu$m/s [S3], the Reynolds number is:
\begin{equation}
\text{Re}_{\text{Vc}} \approx \frac{(1000)(150 \times 10^{-6})(200 \times 10^{-6})}{0.001} \approx 0.03.
\end{equation}

For \textit{V. ferrisii}, with a measured diameter $L \approx 800~\mu$m and a typical swimming speed for large volvocine algae of $v \approx 1$~mm/s [S4], the Reynolds number is:
\begin{equation}
\text{Re}_{\text{Vf}} \approx \frac{(1000)(1 \times 10^{-3})(800 \times 10^{-6})}{0.001} \approx 0.8.
\end{equation}
These values confirm that \textit{V. carteri} operates in the low-Re regime, while \textit{V. ferrisii} operates in the intermediate-Re regime where inertia is significant.

\clearpage
\onecolumngrid

\vspace{1em}
\noindent \textbf{\large Supplemental Figures and Table}
\vspace{1em}

\begin{figure}[h!]
\includegraphics[width=\textwidth]{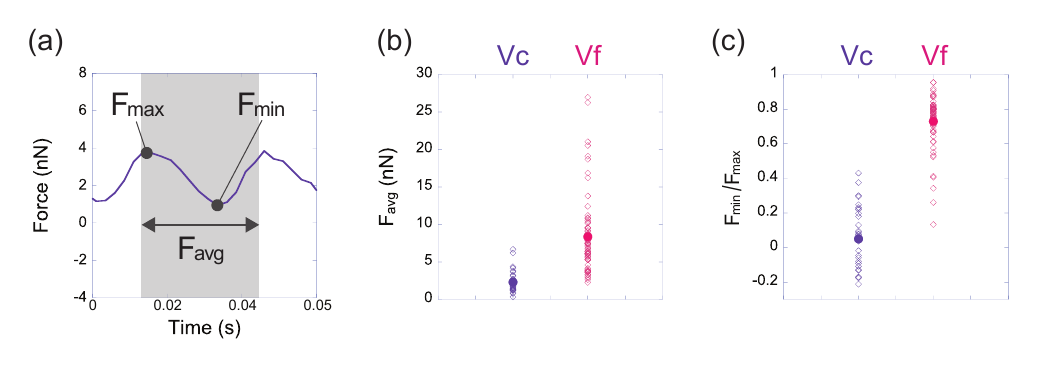}
\caption{\label{fig:quantification} \textbf{Quantification of propulsive force properties.} (a) Schematic defining the key metrics derived from the oscillating force profile: peak force ($F_{\text{max}}$), minimum force ($F_{\text{min}}$), and time-averaged force ($F_{\text{avg}}$). (b) Distribution of the time-averaged propulsive force ($F_{\text{avg}}$) for Vc (blue, n=35) and Vf (orange, n=71). (c) Distribution of the force consistency ratio ($F_{\text{min}}/F_{\text{max}}$) for the same events. The force from Vc is highly pulsatile, while the force from Vf is significantly more continuous.}
\end{figure}

\begin{table}[h!]
\caption{\label{tab:params} \textbf{Physical and mechanical parameters.} Spheroid diameter and force properties are mean values derived from experimental measurements. Spheroid mass and Reynolds number (Re) are calculated from these measurements and literature values for density [S3].}
\begin{ruledtabular}
\begin{tabular}{lcc}
\textrm{Parameter} & \textit{V. carteri} (Vc) & \textit{V. ferrisii} (Vf) \\
\colrule
Spheroid Diameter ($\mu$m) & 200 & 800 \\
Spheroid Mass (ng) & 4.2 & 269 \\
Reynolds Number (Re) & $\sim$0.03 & $\sim$0.8 \\
Peak Force, $F_{\text{max}}$ (nN) & 3.3 & 9.8 \\
Force Consistency ($F_{\text{min}}/F_{\text{max}}$) & 0.05 & 0.73 \\
Oscillation Rate (Hz) & $\sim$30 & $\sim$30 \\
\end{tabular}
\end{ruledtabular}
\end{table}

\end{document}